\documentclass[twocolumn, aps,prx,citeautoscript,superscriptaddress,floatfix,notitlepage]{revtex4-2}
\usepackage[utf8]{inputenc}
\usepackage{natbib}
\usepackage{graphicx}
\usepackage{amsmath}
\usepackage{amssymb,bm,xspace,soul}
\usepackage{color}
\usepackage[dvipsnames]{xcolor}
\usepackage{anyfontsize} %changes font size with \fontsize{size of font}{spacing}\
\usepackage{units}
\usepackage{times}
\usepackage{setspace}
\usepackage{braket}
\usepackage{empheq}
\usepackage{xcolor}
\usepackage[normalem]{ulem}
\usepackage{slashed}
\usepackage{comment}
\usepackage{wasysym}
\usepackage[bbgreekl]{mathbbol}

\DeclareMathAlphabet{\mathcald}{U}{dutchcal}{m}{n}
\SetMathAlphabet{\mathcald}{bold}{U}{dutchcal}{b}{n}
\DeclareMathAlphabet{\mathalt}{U}{dutchcal}{b}{n}

\DeclareFontFamily{U}{BOONDOX-calo}{\skewchar\font=45 }
\DeclareFontShape{U}{BOONDOX-calo}{m}{n}{
  <-> s*[1.05] BOONDOX-r-calo}{}
\DeclareFontShape{U}{BOONDOX-calo}{b}{n}{
  <-> s*[1.05] BOONDOX-b-calo}{}
\DeclareMathAlphabet{\mathcalb}{U}{BOONDOX-calo}{m}{n}
\SetMathAlphabet{\mathcalb}{bold}{U}{BOONDOX-calo}{b}{n}
\DeclareMathAlphabet{\mathbcalbx}{U}{BOONDOX-calo}{b}{n}

\colorlet{linkColour}{magenta}
\colorlet{citeColour}{OliveGreen}
\colorlet{urlColour}{cyan}

\usepackage[colorlinks=true]{hyperref}
\hypersetup{
    unicode=false,          % non-Latin characters
    pdftoolbar=true,        % show Acrobat
    pdfmenubar=true,        % show Acrobat
    pdffitwindow=false,     % window fit to page when opened
    pdfstartview={FitH},    % fits the width of the page to the window
    pdftitle={Fibonacci FQH},    % title
    pdfauthor={Hart Goldman},     % author
    pdfsubject={Subject},   % subject of the document
    pdfcreator={Hart Goldman},   % creator of the document
    pdfproducer={}, % producer of the document
    pdfkeywords={keyword1} {key2} {key3}, % list of keywords
    pdfnewwindow=true,      % links in new window
    colorlinks=true,       % false: boxed links; true: colored links
    linkcolor=linkColour, %red,          % color of internal links (change box color with linkbordercolor)
    citecolor=citeColour,        % color of links to bibliography
    filecolor=linkColour,      % color of file links
    urlcolor=urlColour           % color of external links
}

\allowdisplaybreaks
\newcommand{\sgn}{\operatorname{sgn}}
\newcommand{\Tr}{\operatorname{Tr}}

\newcommand{\para}{$\text{Parafermion}_3$}

\begin{document}

\title{A composite particle construction of the Fibonacci fractional quantum Hall state}
\author{Hart Goldman}
\affiliation{Department of Physics, Massachusetts Institute of Technology, Cambridge, MA 02139}
\author{Ramanjit Sohal}
\author{Eduardo Fradkin}
 \affiliation{Department of Physics and Institute for Condensed Matter Theory, University of Illinois at Urbana-Champaign, Urbana, IL, 61801}
\date{\today}

\begin{abstract}
The Fibonacci topological order is the simplest %example of \textcolor{blue}{of a platform for} 
platform for a universal topological quantum computer, consisting of a single type of non-Abelian anyon, $\tau$, with fusion rule $\tau\times\tau=1+\tau$. While it has been proposed that the anyon spectrum of the $\nu=12/5$ fractional quantum Hall state %contains 
includes a Fibonacci sector, %among others, %\textcolor{blue}{(among others)}, 
a dynamical picture of how a \emph{pure} Fibonacci state may emerge in a quantum Hall system has been lacking. Here we use recently proposed non-Abelian dualities to construct a Fibonacci state of bosons at filling $\nu=2$ starting from a trilayer of integer quantum Hall states. Our parent theory consists of bosonic ``composite vortices" coupled to fluctuating $U(2)$ gauge fields, which is related to the standard theory of Laughlin quasiparticles by duality. The Fibonacci state is obtained by clustering the composite vortices between the layers, %followed by 
along with flux attachment, a  procedure reminiscent of the clustering picture of the Read-Rezayi states. %{\raman \sout{In terms of the fundamental boson degrees of freedom, this procedure binds holes on one layer to charges on the other two, and so we interpret this state as a new type of non-Abelian singlet state.}} %One consequence of this result is that the Fibonacci state is actually a \emph{descendant}. 
We further use this framework to motivate a wave function for the Fibonacci fractional quantum Hall state. 
\end{abstract}

\maketitle
% \tableofcontents

% \section{Introduction}

%\section{Introduction}

%One of the primary impetuses for the realization of non-Abelian topologically ordered phases of matter lies in their potential role 
\emph{Introduction.} Non-Abelian topological orders are among the most promising platforms for %in 
fault-tolerant quantum computation \cite{Nayak2008}. The excitations in these phases are non-Abelian anyons, which are quasiparticles with non-Abelian exchange statistics \cite{Moore1991}. Non-Abelian anyons therefore provide a source of \emph{topological} degeneracy, allowing for non-local storage of information. Information %which 
can then be manipulated through braiding of the anyons, a process which is resilient against decoherence from %classical 
local perturbations because of its topological nature \cite{Chamon-1997,Fradkin-1998,Bonderson-2006,Stern-2006,Bishara-2009}. Among the most promising %platforms to realize 
systems for realizing non-Abelian topological order are %provided by
%two-dimensional 
2d gases of electrons in strong magnetic fields, which can form fractional quantum  Hall (FQH) states. Excitingly, there is mounting experimental evidence for fractional statistics in FQH states \cite{Nakamura-2020}, %and of the presence of
and for a non-Abelian FQH state at filling fraction $\nu=5/2$ %is %in fact 
%a non-Abelian state 
supporting the simplest non-Abelian anyon, the Ising anyon \cite{Radu-2008,Dolev-2008,Willett-2009,Willett-2013}.

Ising anyons, however, are not sufficient for universal %topological
quantum computation \cite{Nayak2008}. In contrast, topological orders supporting the so-called Fibonacci anyon %, $\tau$,  %so-named for its 
%which has a fusion rule $\tau \times \tau = 1+\tau$, 
can serve as universal quantum computers \cite{Freedman-2002}. This follows from the Fibonacci anyon's fusion rule, $\tau\times\tau=1+\tau$, where $\tau$ is the Fibonacci anyon, $1$ is the trivial anyon, and $\times$ denotes anyon fusion.  For this reason, there has been much interest %has arisen in the experimentally 
in the observed $\nu=12/5$ FQH state, as numerics suggest this may correspond to %the (particle-hole conjugate) $k=3$ 
the $\mathbb{Z}_3$ Read-Rezayi (RR) state \cite{Read1999}, which %does support 
supports the Fibonacci anyon among other, Abelian anyons \cite{Pakrouski2016,Mong2017}. % (a possible 2/5 Abelian Jain FQH state has also not yet be excluded). %However, the presence of other, Abelian anyons in the $k=3$ RR state 
Unfortunately, the presence of the other anyons can complicate manipulation of the Fibonacci anyons by entering into braiding processes, a form of quasiparticle poisoning. It is thus of interest to understand if it is possible to %how one could 
realize a topological order supporting the Fibonacci anyon as its \emph{only} excitation.

Several proposals have been put forward for realizing such a Fibonacci state. These include the nucleation of a Fibonacci state on top of an Abelian FQH state using proximity coupled superconductors \cite{Mong2014}, chiral superconducting islands with special couplings \cite{Hu2018},  and the possible realization of the Fibonacci state at an integer filling of Landau levels \cite{Lopes2019}. All these studies follow the spirit of coupled wire constructions \cite{Teo2014} which, although providing concrete and analytically tractable microscopic models %for the emergence of topological order, 
with topologically ordered ground states, do not provide a physical picture for the 
dynamics that could lead to the emergence of such states in an isotropic system. A quantum loop model for a Fibonacci state was proposed in Ref. \cite{Fendley-2005}. In the context of Abelian FQH states, such a picture is provided by composite fermion/boson field theories %provide clear pictures for the emergence of \emph{Abelian} states
\cite{Jain-1989,Zhang-1989,Lopez-1991}. While a composite particle picture is lacking for most non-Abelian states, including %any putative 
the Fibonacci state, %with one
notable exceptions include the Moore-Read FQH state (and its cousins) at $\nu=5/2$, which can be described as arising from the pairing of composite fermions \cite{Read2000}, the Read-Rezayi sequence \cite{Fradkin1999,Goldman2019}, and a range of Blok-Wen states \cite{Radicevic2016,Goldman2020}. Indeed, it is an open problem to establish a precise composite particle picture for any \emph{purely} non-Abelian state, as flux attachment generically leads to Abelian anyon content \footnote{In Ref. \cite{Goldman2020}, we already %used duality to %found that the Fibonacci state can arise at filling $\nu=3$ in a fermionic system
%While we note that a composite particle 
predicted a phase transition between a $\nu=3$ integer quantum Hall (IQH) state of fermions and a Fibonacci state at $\nu=3$, %the same filling, % fraction, %was predicted in our analysis in Ref. \cite{Goldman2020}, 
but this transition involves very strongly interacting physics, and its physical interpretation remains mysterious.}. 

%The goal of this Rapid Communication (?) is to 
In this %Letter
article, we employ recently proposed Chern-Simons-matter field theory dualities \cite{Aharony2012, Giombi2012, Aharony2016} to construct a composite particle theory for the emergence of the Fibonacci state in a QH system of bosons at $\nu=2$, following our earlier approach in Refs. \cite{Goldman2019,Goldman2020}. These dualities can be interpreted as non-Abelian analogues of flux attachment. %We have already employed these dualities in the construction of Landau-Ginzburg theories for the RR and non-Abelian spin singlet states starting from Abelian multilayer parent theories \cite{Goldman2019} and composite fermion theories for the Blok-Wen states \cite{Goldman2020}. 
In the present work, we %aim to 
%provide an alternative description of this state using 
instead use duality to construct a Landau-Ginzburg description of a Fibonacci state of \emph{bosons} starting from a trilayer of IQH states, %with a unit of attached flux that renders
using flux attachment to render the electric charges bosonic. In this setup, the dynamical mechanism leading to the Fibonacci state is manifest as inter-layer clustering of dual bosonic ``composite vortices," which couple to a fluctuating, non-Abelian gauge field.  %Crucially, 
Our chosen clustering mechanism binds electric charges on two of the layers to holes on the third, breaking the inter-layer exchange symmetry. Our flux attachment procedure similarly breaks this symmetry, rendering two of the layers topologically trivial and endowing the remaining layer with the topological order of the Halperin $(2,2,1)$ state.%, leading to a natural interpretation of the Fibonacci state as a spin singlet state {\raman (RS: Do we still want to interpret the construction as describing a spin singlet state?)}, where the inter-layer dipole moment is interpreted as an emergent spin.  %In the balance of this paper, 
%An advantage of 

Our dynamical mechanism therefore has an element of clustering, which underlies the interpretation of the RR states, while retaining the character of a multilayer state, as the $(2,2,1)$ state is commonly interpreted as a bilayer (it has a $\mathbb{Z}_2$ exchange symmetry). %and singlet formation, which motivated the multilayer generalization of the RR states, the non-Abelian spin singlet (NASS) states \cite{Ardonne1999,Ardonne2002}. 
In parallel to this intuition, we motivate an ideal wave function for the Fibonacci state, an as-yet unprecedented achievement. This wave function superficially describes a bilayer state, %(``spin'' up/down) 
but nevertheless has the clustering properties of the $\mathbb{Z}_3$ RR state, which describes clusters of three local quasiparticles. %{\raman electrons \sout{quasiparticles}}.  

%We proceed by first introducing the parent theory in which the Fibonacci state will be realized, a trilayer integer quantum Hall (IQH) state, as well as the non-Abelian duality that we will employ. % will be employed in the Landau-Ginzburg theory. 
%We then use the duality to construct our Landau-Ginzburg theory and, before concluding, ({\raman hopefully}) propose a trial wave function for this state.

\begin{figure}
  \centering
    \includegraphics[width=0.48\textwidth]{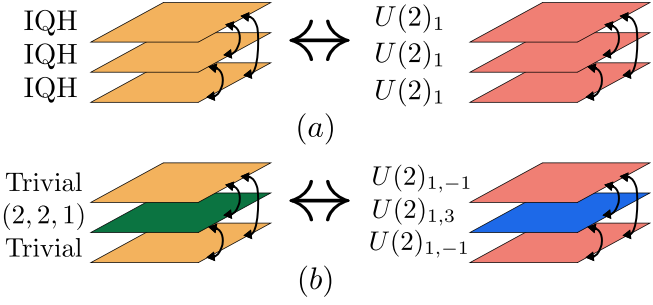}%{stack-alt-v2}
    \caption{A schematic of our construction of LG theories for the (a) $U(2)_3$ and (b) $U(2)_{3,1}$ Fibonacci states. Here $\leftrightarrow$ denotes duality between the parent theories of ordinary Laughlin quasiparticles and dual theories of composite vortices coupled to $U(2)$ gauge fields. The non-Abelian state is obtained by clustering of the dual composite vortices %, which couple to emergent $U(2)$ gauge fields, 
    between the layers. %On the left, we stack 
    %(a) A trilayer of $\nu=2$ IQH states is dual to a trilayer of scalars coupled to separate copies of $U(2)_1$. Interlayer pairing in the latter non-Abelian language%, %indicated by the double-headed arrows, 
    %yields the $U(2)_3$ state. %These correspond to non-local interactions in the original Abelian theory. 
    %(b) Attaching fluxes separately to each layer of (a) yields a system with two trivial layers of scalars and a third layer with a scalar coupled to the Halperin $(2,2,1)$ CS theory. Pairing in a dual non-Abelian theory yields the $U(2)_{3,1}$ state, which describes the Fibonacci topological order.} 
    }
    \label{fig:cartoon}
\end{figure}

\emph{Parent model and non-Abelian duality.} Our starting point is a trilayer of $\nu=2$ IQH states, as shown in Fig. \ref{fig:cartoon}. %{\hart (add figure)}. 
We will take each layer layer to be near a $\nu=2\rightarrow1$ transition described by a free Dirac fermion in the clean limit,
\begin{align}
\label{eq: parent IQH}
\mathcal{L}_{\mathrm{IQH}}&=\sum_{n=1}^3\left[\bar\Psi_n(i\slashed{D}_A-M)\Psi_n-\frac{3}{2}\frac{1}{4\pi}AdA\right]\,.
\end{align}
Here $\Psi_n$ is a two-component Dirac fermion on layer $n$ \footnote{Here we approximate the Atiyah-Patodi-Singer $\eta$-invariant as a level-$1/2$ Chern-Simons term and include it in the Lagrangian.}, $A_\mu$ is the background electromagnetic (EM) gauge field, and we use the notation $D_B^\mu=\partial^\mu-iB^\mu$, $BdC=\varepsilon^{\mu\nu\lambda}B_\mu\partial_\nu C_\lambda$, and $\slashed{B}=B^\mu\gamma_\mu$, where $\gamma^\mu$ are the Dirac gamma matrices.  Integrating out the Dirac fermions %reveals that the theory is in 
yields a $\nu=2$ ($\nu=1$) IQH phase for $\sgn(M)<0$ ($\sgn(M)>0$). %and a $\nu=1$ IQH phase for $\sgn(M)>0$. 
Note %that 
we define the filling as $\nu=-2\pi\rho_e/B$, $\rho_e=\langle\delta\mathcal{L}/\delta A_0\rangle,B=\varepsilon^{ij}\partial_iA_j$. Our interest will be in the physics near the quantum phase transition at $M=0$.

Near $M=0$, this theory has been proposed to satisfy a large number of boson-fermion dualities \cite{Aharony2016}, which are relativistic generalizations of the familiar flux attachment duality that relates the IQH transition of fermions to the condensation of composite bosons \cite{Zhang-1989}.  These relate the free Dirac fermion theory on each layer to one of a Wilson-Fisher boson, $\phi_n$, coupled to a fluctuating $U(N)$ Chern-Simons (CS) gauge field, $a_n$, in the fundamental representation \cite{Hsin2016,Chen2018,Hui2017}. While a free Dirac fermion has a bosonic dual for any value of $N$, our interest will be in the case of $N=2$,
\begin{align}
\label{eq: bosonic parent}
\tilde{\mathcal{L}}_{\mathrm{IQH}}=&\sum_{n=1}^3\left[|D_a\phi_n|^2-r|\phi_n|^2-|\phi_n|^4\right]\nonumber\\&+\sum_{n=1}^3\mathcal{L}_{\mathrm{CS}}[a_n]+A_\mu j_{\mathrm{top}}^\mu\,,\\
\mathcal{L}_{\mathrm{CS}}[a_n]=&\,\frac{1}{4\pi}\Tr\left[a_nda_n-\frac{2i}{3}(a_n)^3\right]\,,\\
\label{eq: jtop}
A_\mu j^\mu_{\mathrm{top}}=&\,\frac{1}{2\pi}Ad\Tr[a_1-a_2+a_3]\,.
\end{align}
Here $-|\phi|^4$ denotes tuning such that the Wilson-Fisher fixed point occurs at $r=0$, and traces are over color (i.e. $U(2)$) indices. We have also selected the BF terms in Eq. \eqref{eq: jtop} such that the second layer has opposite EM charge from the other two. Because each layer is decoupled from one another, we may freely determine the signs in Eq. \eqref{eq: jtop} because the partition function has a charge conjugation symmetry. 

The fact that the theory in Eq. \eqref{eq: bosonic parent} has the same phase diagram as that of Eq. \eqref{eq: parent IQH} follows from the so-called level-rank duality of topological quantum field theories (TQFTs) \cite{Camperi-1990,Naculich-1990,Naculich-1990b,Hsin2016}, which is an equivalence between $U(N)_k$ and $SU(k)_{-N}$ CS theories, where the subscript is the CS level. In particular, one can set $k=1$, leading to a duality between a trivial (i.e. IQH) theory and a $U(N)_1$ CS theory,
\begin{align}
\label{eq: SU(1) lvl-rk}
\mathcal{L}_{\mathrm{CS}}[b]+\frac{1}{2\pi}Ad\Tr[b]&\longleftrightarrow-\frac{N}{4\pi}AdA\,,
\end{align}
where $b$ is a $U(N)$ gauge field, and we have suppressed gravitational Chern-Simons terms. 

Using level-rank duality, we can check the phase diagram of Eq. \eqref{eq: bosonic parent}: for $\sgn(r)>0$, the $\phi$ bosons are gapped, leading to a $U(2)_1$ theory on each layer, which describes a trilayer of $\nu=2$ IQH states by Eq. \eqref{eq: SU(1) lvl-rk}. Similarly, for $r<0$ the bosons condense, breaking the gauge group down to $U(1)$ on each layer. Integrating out the remaining $U(1)$ gauge fields leads to the desired trilayer $\nu=1$ response. The equivalence of the phase diagrams of the theories in Eqs. \eqref{eq: parent IQH} and \eqref{eq: bosonic parent} has led to the conjecture that the critical points at $r=M=0$ are identical. Below we will assume this to be the case, our confidence bolstered by the large-$N,k$ derivations of Refs. \cite{Giombi2012,Aharony2012} and the Euclidean lattice derivation of Ref. \cite{Chen2018}. 

\emph{Landau-Ginzburg theory.} To target the Fibonacci phase, we first identify a CS TQFT representation of the state. It was recently shown \cite{Cordova2018} that one such representation is %$U(2)_{3,1}$ \cite{Cordova2018}, %defined as %\textcolor{blue}{where}
\begin{align}
	U(2)_{3,1} &= \frac{SU(2)_3 \times U(1)_2}{\mathbb{Z}_2}. \label{eq: U(2)_3,1}
\end{align}
This is a $U(2)$ CS gauge theory where the Abelian and non-Abelian parts of the gauge field have different CS levels. The quotient by $\mathbb{Z}_2$ simply enforces that these two components are part of the same $U(2)$ gauge field {\footnote{ {  See %Supplemental Material at [URL will be inserted by publisher] 
Appendix for our CS conventions and technical details involved in our effective field theory and wave function constructions.}} }%(see Supplemental Material)
. The Lagrangian for this theory is written as
\begin{align}
\label{eq: Fib Lagrangian}
\mathcal{L}_{\mathrm{Fib}}&=3\,\mathcal{L}_{\mathrm{CS}}[a]-\frac{1}{4\pi}\Tr[a]\,d\Tr[a]+\frac{1}{2\pi}Ad\Tr[a]\,,
\end{align}
where $a$ is again a $U(2)$ gauge field. One can check that this theory has a single nontrivial anyon, besides the vacuum, which transforms in the spin-1 representation of $U(2)$, satisfies the Fibonacci fusion rule, $\tau\times\tau=1+\tau$, and has topological spin $h_\tau=2/5$ \cite{Note3}. %we make use of another, less trivial level-rank duality. 
We also comment that this theory is known to be dual to a $(G_2)_1$ TQFT, where $G_2$ is the smallest exceptional simple Lie group \cite{Bonderson2007,Mong2017,Cordova2018}.    

To access the $U(2)_{3,1}$ state, we start by introducing inter-layer clustering to the composite vortex theory, Eq. \eqref{eq: bosonic parent}, via coupling to a scalar field, $\Sigma_{nm}$,
\begin{align}
\label{eq: clustering term}
\mathcal{L}_{\mathrm{cluster}}&=-\sum_{n,m}\phi^\dagger_m\Sigma_{mn}\phi_n-V[\Sigma]\,.
\end{align}
Under gauge transformations, $\Sigma_{nm}\mapsto U_m\Sigma_{mn}U^\dagger_n$, where $U_n$ is a $U(2)$ gauge transformation on layer $n$. It can be understood as a Hubbard-Stratonovich field associated with the order parameter, $\phi^\dagger_m\phi_n$. We choose the potential $V[\Sigma]$ such that
\begin{align}
\label{eq: sigma vev}
\langle \Sigma_{mn}\rangle&=
M_{mn}\,\mathbf{1}_{2}\,,M_{mn}\neq0\,,\,M_{nn}>0\,,\det M>0\,,
\end{align}
where $\mathbf{1}_{2}$ is the $2\times 2$ identity matrix in color space and $M_{mn}$ is a constant Hermitian matrix. In the resulting ground state, the $\phi_n$ fields are individually gapped, while the clustering order parameter, $\phi^\dagger_m\phi_n$ is condensed. 

Because Eq. \eqref{eq: sigma vev} is invariant under gauge transformations where $U_1=U_2=U_3$, the gauge group is broken as $U(2)\times U(2)\times U(2)\rightarrow U(2)$, Higgsing gauge field configurations except for those with $a_1=a_2=a_3\equiv a$. As a result, the CS terms for each of the $a_n$ gauge fields add, leading to a $U(2)_3$ theory, 
\begin{align}
\label{eq: U(2)3}
\mathcal{L}_{U(3)_3}[a,A]=3\,\mathcal{L}_{\mathrm{CS}}[a]+\frac{1}{2\pi}Ad\Tr[a]\,.
\end{align}
Computing the Hall response by integrating out $\Tr[a]=\Tr[\tilde{a}\mathbf{1}_2]=2\tilde{a}$, one finds that the total filling fraction is now $\nu=2/3$, rather than $\nu=6$. The change in the filling fraction is related to our choice of charge assignments in Eq. \eqref{eq: jtop}, which results in the unit coefficient of the BF term in Eq. \eqref{eq: U(2)3}. While in the decoupled trilayer theory this choice of signs was immaterial, upon clustering the EM charge densities on each layer, $\rho_n=\varepsilon_{ij}\partial^i\Tr[a_n^j]/2\pi$, $i,j=x,y$, are pinned as $\rho_1=\rho_3=-\rho_2$, thereby breaking the discrete symmetry exchanging the layers and altering the filling fraction. The resulting minimal EM charge will prove crucial to %ultimately 
obtaining the Fibonacci state. %Because the density of electric charges on the first and third layers is the density of holes on the second layer, the charged excitations have vanishing inter-layer dipole moment, and so we refer to them as singlets.

The Fibonacci state, Eq. \eqref{eq: Fib Lagrangian}, is a \emph{descendant} of the $U(2)_3$ state at $\nu=2/3$. To see this, we attach a single unit of flux to the ``electrons," the charges which couple to the background EM vector potential, $A_\mu$, and are understood to be the vortices of $\Tr[a]$ in the variables of Eq. \eqref{eq: U(2)3}. Since in our starting theory, Eq. \eqref{eq: parent IQH}, the EM charges are fermions, flux attachment shifts their statistics and renders the fundamental EM charges bosonic. Explicitly, introducing an Abelian statistical gauge field, $b$, we have
\begin{align}
\label{eq: flux attachment}
\mathcal{L}&=\mathcal{L}_{U(3)_3}[a,b]+\frac{1}{4\pi}bdb+\frac{1}{2\pi}bdA+\frac{1}{4\pi}AdA\,.
\end{align}
Integrating out $b$, one immediately finds the Lagrangian in Eq. \eqref{eq: Fib Lagrangian}, which displays a $\nu=2$ Hall response. We have therefore found, using a combination of flux attachment and inter-layer clustering, a Fibonacci state of bosons at $\nu=2$. 

The flux attachment transformation in Eq. \eqref{eq: flux attachment} transmutes the original electric charges, which are fermions, to bosons, but it also mixes the three layers of the parent model, Eq. \eqref{eq: parent IQH}. %To glean more physical insight, one may instead wish to construct a parent trilayer state with bosonic charges on each layer. 
A more physically transparent approach, which also leads to a Fibonacci state at $\nu=2$, % It is possible to show (see Supplemental Material) that the process of clustering followed by flux attachment as in Eq. \eqref{eq: flux attachment} can be equivalently implemented 
proceeds by first attaching a positive flux to each electron on the first and third layers of the theory in Eq. \eqref{eq: parent IQH} while attaching a negative flux to each electron on the second layer, explicitly breaking the layer exchange symmetry outright and leading to the parent theory depicted in Fig. \ref{fig:cartoon}(b). On the first and third layers, this results in theories of electrically charged Wilson-Fisher bosons %(i.e. they do not couple to statistical gauge fields) 
on top of a $\nu=-2$ IQH state. On the second layer, however, this leads to Wilson-Fisher bosons coupled to %\textcolor{blue}{\sout{a Halperin $(2,2,1)$} the} 
the Halperin $(2,2,1)$ CS gauge theory %\textcolor{blue}{associated with the Halperin $(2,2,1)$ bosonic FQH state at} 
at filling $\nu=+2/3$. %({\hart I will add equations if there is room}). 
One can show %{\raman \sout{(see Supplementary Material)}}
that clustering of composite vortices starting from this trilayer state leads to a Fibonacci FQH state \cite{Note3}.
%In other words, if flux attachment is carried out prior to clustering, the parent theory Eq. \eqref{eq: parent IQH} becomes the bosonic trilayer state depicted in Fig. \ref{fig:cartoon}(b).% ({\hart include figure}).        
We note that the Halperin $(2,2,1)$ state has appeared as a parent state for the Fibonacci order in related constructions \cite{Mong2014,Clarke2015}.

\begin{figure}
\includegraphics[width=0.4\textwidth]{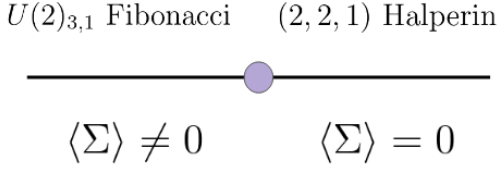}
\caption{Phase diagram in terms of the clustering order parameter, $\Sigma$. The ordered state,  $\langle\Sigma_{mn}\rangle=M_{mn}\mathbf{1}_2$, corresponds to the bosonic Fibonacci FQH state at filling $\nu=2$. The disordered phase, $\langle\Sigma_{mn}\rangle=0$, is a decoupled trilayer with the topological order of the Halperin $(2,2,1)$ state at total filling $\nu=-4+2/3=-10/3$.}
\label{fig: phase diagram}
\end{figure}

Using this bosonic parent description of Fig. \ref{fig:cartoon}(b), the final Landau-Ginzburg theory of the Fibonacci state can be expressed in terms of the clustering order parameter, $\Sigma$, after integrating out the composite vortices, $\phi$, and the auxiliary gauge fields associated with flux attachment,% (see Supplementary Material),
\begin{align}
\mathcal{L}=&\sum_{m,n}\Tr\left[|\partial\Sigma_{mn}-ia_m\Sigma_{mn}+i\Sigma_{mn} a_n|^2\right]+\sum_n\mathcal{L}_{\mathrm{CS}}[a_n]\nonumber\\
%&+\sum_n\mathcal{L}_{\mathrm{CS}}[a_n]+b_\mu j^\mu_{\mathrm{top}}+\frac{1}{4\pi}(b+A)d(b+A)\,,
&+\sum_n(-1)^n\left(\frac{1}{4\pi}\Tr[a_n]d\Tr[a_n]+\frac{1}{2\pi}Ad\Tr[a_n]\right)\nonumber\\
&-V_r[\Sigma]\,.
\end{align}
where the first term is a kinetic term generated by quantum corrections due to integrating out $\phi$, and $V_r$ is the renormalized potential for $\Sigma$. The trace is again over color indices. The phase diagram can be understood as in Fig. \ref{fig: phase diagram}. For $\langle\Sigma\rangle=0$, the theory consists of three decoupled layers: two IQH insulators and a single Halperin $(2,2,1)$ layer. For $\langle\Sigma\rangle=M\neq0$, the theory finds itself in a %non-Abelian excitonic insulator 
phase 
with Fibonacci topological order.  
%{\hart (Note to self: Insert discussion of the phase diagram -- HG)}
%\textcolor{blue}{Is Fig. 1 a phase diagram?}

Furthermore, one can identify the Fibonacci anyons with gapped degrees of freedom in the Landau-Ginzburg theory; namely, the excitations of the adjoint bilinear of composite vortices, $\phi^\dagger t^a\phi$, where $t^a$ are the generators of $SU(2)\subset U(2)$. This can be observed from the fact that this operator transforms in the spin-1 representation of the gauge group and has vanishing electric charge, both properties of the Fibonacci anyon. Note that while the $\phi$ fields possess a layer index, in the Fibonacci state this does not lead to any unwanted degeneracy due to the condensation of $\langle\phi^\dagger_m\phi_n\rangle$, and so there is only one Fibonacci anyon.   

\emph{Fibonacci wave function.} 
%Although our 
Having developed an effective field theory that provides a concrete dynamical mechanism for how the Fibonacci state may be realized in a bosonic system at $\nu=2$, %it would be useful to also have a trial wave function for this exotic state. 
we now seek to develop an ideal wave function, which until now has also proven elusive. %A mainstay in the study of FQH states, trial 
Ideal wave functions encode information about the clustering properties of electrons in non-Abelian states and can be compared with numerically obtained ground states in order to identify the topological order realized in realistic Hamiltonians. Remarkably, the wave function we will obtain displays a number of physical features that parallel the above effective field theory construction. 
%With this in mind, and with the goal of supplementing the intuition gained from our field theory construction, we employ the standard conformal field theory (CFT) approach to set up an ideal wave function for the Fibonacci state.
%Our effective field theory provides a concrete dynamical picture for how the Fibonacci state may be realized in a bosonic system at $\nu=2$. A complementary approach to describing FQH states is provided by the construction of trial wave functions. These encode information about the clustering properties of electrons in non-Abelian states and can be compared with numerically obtained ground states in order to identify the topological order realized in realistic Hamiltonians. %Indeed, it would be of great interest to understand the competition of the Fibonacci state with the bosonic IQH state at $\nu=2$ \cite{Senthil-2013}. 
%To that end, we employ the standard conformal field theory (CFT) approach to set up an ideal wave function as a supplement to our field theory construction.

%Indeed, as is now well established \cite{Moore1991}, ideal wave functions for FQH states can be expressed in terms of correlators of a CFT: 
To obtain a wave function, we employ the standard conformal field theory (CFT) approach, in which the wave function is constructed in terms of correlation functions of the edge $(G_2)_1 \cong U(2)_{3,1}$ Wess-Zumino-Witten (WZW) CFT, %Following the %standard 
%construction of Ref. \cite{Moore1991}, we expect to be able to express the Fibonacci wave function in terms of a correlation function in the $(G_2)_1 \cong U(2)_{3,1}$ Wess-Zumino-Witten CFT:
$\Psi(\{z_i^\sigma\}) = \langle \prod_{i=1}^N \Psi_\sigma(z_i^\sigma) \rangle$ \cite{Moore1991}. 
Here, $z_i^\sigma = x_i^\sigma + iy_i^\sigma$ are the complex coordinates of the electrons, $\sigma=1,\dots,n_f$ a type of %possible
``flavor" index, $n_f N$ is the number of electrons, and $\Psi_\sigma(z_i)$ are operators in the CFT.
Physically, $\Psi_\sigma(z)$ represents an electron operator and can in general be written as the product $\Psi_\sigma(z) = \psi_\sigma(z) e^{i\varphi (z) / \sqrt{\nu}}$, where $\nu$ is the filling fraction and $\varphi$ is a compact boson. The $\psi_\sigma(z)$ operators are electrically neutral.
%For a FQH state described by a CS theory with gauge group $\mathcal{G}$ at level $k$, the CFT of interest in writing down the ideal wave function is the $\mathcal{G}_k$ Wess-Zumino-Witten CFT. 
%For the Fibonacci FQH state, the CFT of interest in writing down the ideal wave function is the $(G_2)_1 \cong U(2)_{3,1}$ Wess-Zumino-Witten CFT.
From Eq. \eqref{eq: U(2)_3,1}, we %see 
observe that for the case at hand the $\psi_\sigma$'s are %will be
operators in the $SU(2)_3$ CFT, and $e^{i\varphi/\sqrt{\nu}}$, with $\nu=2$, is an operator in the $U(1)_2$ CFT. 

The first step in constructing a wave function is therefore to determine the electron operators, $\Psi_\sigma$. %Our claim is that the 
We claim that the appropriate choice of electron operators is %given by
\begin{align}
\label{eq: electron operators}
	\Psi_\uparrow \equiv \psi_2 e^{i\phi / \sqrt{6} + i\varphi/\sqrt{2}} , \quad \Psi_\downarrow \equiv \psi_1 e^{-i\phi / \sqrt{6} + i\varphi/\sqrt{2}}.
\end{align}
Here we have made use of the fact that operators in the $SU(2)_3$ CFT can be expressed as products of vertex operators of another compact boson, $\phi$, and so-called $\mathbb{Z}_3$ parafermions \cite{Zamoldchikov1985}, $\psi_{1}$ and $\psi_2$, which satisfy the operator product expansions (OPEs),
\begin{align}
    %\begin{split}
	\notag\psi_1(z) \psi_1(z') &\sim (z-z')^{-2/3} \psi_2 (z') + \dots \, \text{(likewise for } 1\leftrightarrow 2) \\
	\psi_1(z) \psi_2(z') &\sim (z-z')^{-4/3} + \dots \, . \label{eq: parafermion-ope}
	%\end{split} 
\end{align}
%which are summarized in the fusion rules $\psi_1 \times \psi_1 = \psi_2$, $\psi_1 \times \psi_2 = 1$. Now, performing the $\mathbb{Z}_3$ quotient in Eq. \eqref{eq: SU(2)_3} amounts to extending the current algebra by Abelian operators in the product theory with conformal spin unity (i.e. bosonic excitations) and satisfying $\mathbb{Z}_3$ fusion rules. 
The choice of the \emph{two} electron operators (labeled by ``spin" $\uparrow/\downarrow$) in Eq. \eqref{eq: electron operators} is motivated by the effective field theory construction discussed above. Indeed, the $(2,2,1)$ Halperin state involved in the parent state in Fig. \ref{fig:cartoon}(b) has two species of vortices satisfying a $\mathbb{Z}_2$ exchange symmetry and is commonly understood as a bilayer state; %It is therefore is commonly understood as a bilayer; 
the remaining two layers in Fig. \ref{fig:cartoon}(b) are topologically trivial. We therefore anticipate that the Fibonacci wave function ``knows" about this exchange symmetry and choose electron operators as such.

%Note that, motivated by our field theory, we have defined \emph{two} species of electron operators (labeled by ``spin" $\uparrow/\downarrow$). 
%Aside from this formal construction, the presence of two electron species can be motivated from our field theory. 
%Indeed, although our Landau-Ginzburg theory of the Fibonacci state made use of clustering between three layers, two of these layers are topologically trivial, while the third has the topological content of the $(2,2,1)$ Halperin state, which naturally describes a \emph{bilayer} system. 
%At a formal level, 
More formally, the need for two electron species arises from the fact that the electron operators must correspond to 
%The formal reason for this is that the $\Psi^\sigma$ operators defined above correspond to 
generators of the $(G_2)_1$ current algebra, all of which represent local excitations. These can be labeled by the twelve roots of $G_2$, of which two are linearly independent. This suggests that we should have two distinct electron operators, %. This suggests that the Fibonacci FQH state most naturally appears in a two-flavor system, 
as is the case for other FQH wave functions based on rank-two Lie algebras \cite{Ardonne1999,Ardonne2002,Barkeshli2010}. 
%The construction of this chiral algebra is rather technical, and so is relegated to the Supplemental Material. 
Following Refs. \cite{Ardonne1999,Ardonne2002}, we require that our choice of electron operators %is dictated by the requirements 
is such that they have the same electric charge %, represented by the fact that they both contain $e^{i\varphi/\sqrt{2}}$ factors, 
and opposite $SU(2)$ spin. The first requirement is satisfied via the two $e^{i\varphi/\sqrt{2}}$ factors; the second %latter %represented 
by the fact that their $SU(2)_3$ factors are conjugate to one another %. 
%The technical details of this construction are presented in the Appendix %relegated to the Supplemental Material 
\cite{Note3}. 

%With this identification of the electron operators, %we can then write the Fibonacci wave function as
%\begin{align}
%	\Psi(\{ z_i, w_i \} ) &= \langle \prod_{i=1}^N \Psi_\uparrow(z_i) \Psi_\downarrow (w_i) O_{bg} \rangle %\\
%(w_i)} O_{bg} \rangle,
%\end{align}
%We can now proceed to writing 
The Fibonacci wave function can thus be written as a $2N$-point correlation function of the $\Psi_{\uparrow / \downarrow}$ operators. The correlators of the vertex operators can be explicitly evaluated, and so we obtain (%dropping the usual
up to an overall Gaussian factor),
% Here $O_{bg}$ is a background charge operator that ensures the correlator of the $\varphi$ vertex operators is non-vanishing. We can evaluate the correlators of the vertex operators to obtain (dropping an overall Gaussian factor)
\begin{align}
    \begin{split}
	\Psi(\{ z_i, w_i \} ) %&= \langle \prod_{i=1}^N \psi_2(z_i) \psi_1(w_i) \rangle \langle \prod_{i=1}^N e^{i \frac{1}{\sqrt{6}}\phi (z_i)} e^{-i \frac{1}{\sqrt{6}}\phi (w_i)} \rangle \langle \prod_{i=1}^N e^{i \frac{1}{\sqrt{2}}\varphi (z_i)} e^{i \frac{1}{\sqrt{2}}\varphi (w_i)} O_{bg} \rangle \\
	&= \langle \prod_{i=1}^N \psi_2(z_i) \psi_1(w_i) \rangle  \prod_{i,j} (z_i - w_i)^{1/3} \\
	& \qquad \times \prod_{i<j} (z_i - z_j)^{2/3} \prod_{i<j} (w_i - w_j)^{2/3}, \label{eq: fibonacci-wave function}
	\end{split}
\end{align}
where $z_i$ ($w_i$) labels the position of the up (down) ``spin." %and $w_i$ label the positions of the up and down ``spins".
This formal expression encodes key properties of the Fibonacci state. Indeed, the highest power of $z_1$ appearing in the factors multiplying the parafermion correlator is $2N(1/2)$, 
yielding a filling fraction of $\nu=2$, consistent with our field theory construction. 
Additionally, one can use Eq. \eqref{eq: parafermion-ope} to see that the wave function satisfies the same three-body clustering  as the $\mathbb{Z}_3$ RR wave function \cite{Read1999} separately in each of the $z_i$ and $w_i$ coordinates, dovetailing with our description in terms of clustering of composite vortices. These parallels between our proposed wave function and our %effective field theory 
dynamical construction above are encouraging, giving us confidence that  Eq. \eqref{eq: fibonacci-wave function} does indeed describe the Fibonacci state. %This nicely matches the trilayer clustering employed in our field theory construction. 
%We note, however, that the wave function exhibits a simple pole as we bring $z_i \to w_i$. We expect that this short-distance singularity can be regularized without altering the topological properties of the wave function.   
%Moreover, 

By using Eq. \eqref{eq: parafermion-ope} to %effectively 
point-split $\psi_2$ into a product of $\psi_1$'s, one can explicitly evaluate the above parafermion correlator to express Eq. \eqref{eq: fibonacci-wave function} as
\begin{align}
	\Psi(\{ z_i, w_i \} )&= \frac{\Psi_{RR}^{k=3}(\{ z_i , z_i , w_i\} )}{ \prod_{i<j} (z_i - z_j)^{2} \prod_{i,j} (z_i-w_j)},
\end{align}
where $\Psi_{RR}^{k=3}(\{ z_i , z_i , w_i\} )$ is the bosonic $\nu=3/2$ RR wave function for $3N$ particles, with the coordinates of $N$ pairs of particles set equal to one another \cite{Note3}. The apparent asymmetry in $z_i$ and $w_i$ is an artifact of choosing to point-split the $\psi_2$'s. A manifestly symmetric wave function can be obtained %by combining the above expression with that 
via symmetric combination with the wave function obtained by point-splitting the $\psi_1$'s. %We note, however, that 
Note that while the wave function exhibits a simple pole as we bring $z_i \to w_i$, %. We 
we expect that this short-distance singularity can be regularized without altering the topological properties of the wave function.   
%The parallels between our proposed wave function and our effective field theory are encouraging, giving us confidence that  Eq. \eqref{eq: fibonacci-wave function} does indeed describe the Fibonacci state. %, suggesting that Eq. \eqref{eq: fibonacci-wave function} does describe the Fibonacci state and could serve as a basis for numerical investigations for its emergence. 

\emph{Discussion.}
In this %letter 
article, we have presented both a field-theoretic construction of the bosonic Fibonacci state at $\nu=2$ based on non-Abelian composite particle dualities, as well as an explicit wave function for this state. Our construction involves a parent trilayer system, in which the Fibonacci state is realized via clustering of dual ``composite vortices" coupled to fluctuating $U(2)$ gauge fields. Leveraging this construction, we obtain a wave function for the Fibonacci state sharing many of the physical properties of our field-theoretic construction. %which undergoes a clustering transition to the non-abelian state. 
%The relative simplicity of this construction vis-\`a-vis earlier proposals 
Our approach can therefore be used to generate many other exotic states in need of a microscopic construction, as well as to motivate their wave functions.

Unlike other non-Abelian states, short-distance constructions of the Fibonacci state, particularly in isotropic systems, have proven elusive. The fact that our construction is based on a parent state involving fairly germane bosonic FQH phases suggests that a Fibonacci state may be realizable in the laboratory. %Indeed, because our construction pins the electron densities of two layers with the hole density of a third, this suggests that perhaps a transition to a Fibonacci state may be tuned via repulsive interactions. 
Furthermore, the fact that the wave function for the $\nu=2$ bosonic Fibonacci state is manifestly holomorphic clearly suggests that it should be the ground state of a local Hamiltonian projected into a Landau level, and we hope that our wave function will motivate numerical studies in this direction.
Additionally, going forward, %there are two interesting questions that need to be addressed. One is to 
it will be of interest to construct a transparent fermionic analogue of the bosonic Fibonacci state presented here, which would reproduce the state found in Ref. \cite{Goldman2020}. %Also, the fact that the wave function for the $\nu=2$ bosonic Fibonacci state is manifestly holomorphic clearly suggests that it should be the ground state of a local Hamiltonian projected into a Landau level. Additionally, 
%Additionally, 

One may ask whether a different choice of electron operators would have yielded an equally reasonable candidate wave function. In particular, the $\Psi_{\uparrow/\downarrow}$ operators we defined are part of an $SU(2)$ quartet.
%{\raman (This is part is perhaps too speculative)} 
%As described above, writing down the wave function required making a choice of electron operators among the generators of the generators of $(G_2)_1$. In fact, 
%{\raman (To be integrated into discussion)}
%Finally, one may ask whether a different choice of electron operators would have yielded an equally reasonable candidate wave function. In particular, the $\Psi_{\uparrow/\downarrow}$ operators we defined are part of an $SU(2)$ quartet. %, and so one may ask whether the other pair of operators within this quartet constitute an equally viable choice of electron operators. Interestingly, 
For example, the wave function one obtains by choosing the other pair of operators within this quartet as the electrons describes the \emph{Abelian} Halperin $(2,2,-1)$ state.
%On one hand, 
%While this gives us some confidence that our choice of electron operators %were 
%is the correct ones to make in order to obtain the non-Abelian Fibonacci wave function. On the other hand, this may suggest that the $(2,2,-1)$ state is a natural competitor of the Fibonacci state at $\nu=2$.
While it is possible to obtain this state from our parent trilayer theory, it would be interesting to explore how different choices of electron operator in the CFT language may represent different parts of the bulk phase diagram. %It would be interesting to see if our field theory construction allows for accessing this (abelian) state.

\emph{Acknowledgements.} We thank J. Alicea, B. Han, S. Raghu, S. Simon, T. Senthil, M. Stone, J.C.Y. Teo, and C. Xu for discussions and comments on the manuscript. This work was supported in part by the Gordon and Betty Moore Foundation EPiQS Initiative through Grant No. GBMF8684 at the Massachusetts Institute of Technology (HG), by the Natural Sciences and Engineering Research Council of Canada (NSERC) [funding reference number 6799-516762-2018] (RS), and by the US National Science Foundation grant DMR 1725401 at the University of Illinois (EF).

%\begin{appendix}
%\onecolumngrid
%\section{}
%\end{appendix}

\bibliography{Fibonacci}

%\newpage
\clearpage

\onecolumngrid

%\section*{Supplemental Material}
\section*{Appendix}

\subsection{Chern-Simons conventions \label{sec: SM-CS-conventions}}

Here we lay out our conventions for non-Abelian Chern-Simons gauge theories. We define $U(N)$ gauge fields $a_\mu=a^b_\mu t^b$, where $t^b$ are the (Hermitian) generators of the Lie algebra of $U(N)$, which satisfy $[t^a,t^b]=if^{abc}t^c$, where $f^{abc}$ are the structure constants of $U(N)$. The generators are normalized so that $\Tr[t^bt^c]=\frac{1}{2}\delta^{bc}$. The trace of $a$ is a $U(1)$ gauge field, which we require to satisfy the Dirac quantization condition,
\begin{equation}
\label{eq: flux quantization}
\int_{\Sigma} \frac{d\Tr[a]}{2\pi}=n\in\mathbb{Z}\,.
\end{equation}
where $\Sigma\subset X$ is an oriented 2-cycle in spacetime, which we denote $X$. If $a_\mu$ couples to fermions, then it is a spin$_c$ connection, and it satisfies a modified flux quantization condition
\begin{equation}
\int_\Sigma\frac{d\Tr[a]}{2\pi}=\int_\Sigma\frac{w_2}{2}+n\,,\,n\in\mathbb{Z}\,,
\end{equation}
where $w_2$ is the second Stiefel-Whitney class of $X$. In general, the Chern-Simons levels for the $SU(N)$ and $U(1)$ components of $a$ can be different. We therefore adopt the standard notation \cite{Aharony2016},
\begin{equation}
U(N)_{k,k'}=\frac{SU(N)_k\times U(1)_{Nk'}}{\mathbb{Z}_N}\,.
\end{equation}
By taking the quotient with $\mathbb{Z}_N$, we are restricting the difference of the $SU(N)$ and $U(1)$ levels to be an integer multiple of $N$,
\begin{equation}
k'=k+nN\,,n\in\mathbb{Z}\,.
\end{equation}
This enables us to glue the $U(1)$ and $SU(N)$ gauge fields together to form a gauge invariant theory of a single $U(N)$ gauge field $a=a_{SU(N)}+\tilde{a}\,\mathbf{1}$, with $\Tr[a]=N\tilde{a}$ having quantized fluxes as in Eq. \eqref{eq: flux quantization}. The Lagrangian for the $U(N)_{k,k'}$ theory can be written as
\begin{align}
\mathcal{L}_{U(N)_{k,k'}}&=\frac{k}{4\pi}\Tr[ada-\frac{2i}{3}a^3]+\frac{k'-k}{4\pi N}\Tr[a]d\Tr[a]\\
&=\frac{k}{4\pi}\Tr\left[a_{SU(N)}da_{SU(N)}-\frac{2i}{3}a_{SU(N)}^3\right]+\frac{Nk'}{4\pi}\tilde{a}d\tilde{a}\,.
\end{align}
For the case $k=k'$, we simply refer to the theory as $U(N)_k$. 

Throughout this paper, we implicitly regulate non-Abelian (Abelian) gauge theories using Yang-Mills (Maxwell) terms, as opposed to dimensional regularization \cite{Witten1989,Chen1992}. In Yang-Mills regularization, there is a one-loop exact shift of the $SU(N)$  level, $k\rightarrow k+\operatorname{sgn}(k)N$, that does not appear in dimensional regularization. Consequently, to describe the same theory in dimensional regularization, one must start with a $SU(N)$ level $k_{\mathrm{DR}}=k+\operatorname{sgn}(k)N$. The dualities discussed in this paper  %e.g. Eqs. \eqref{eq: U/SU}-\eqref{eq: U/U}, 
therefore would take a somewhat different form in dimensional regularization.

\subsection{Derivation of the bosonic parent state from intra-layer flux attachment}

Here we describe the intra-layer flux attachment procedure described in the main text, which yields the bosonic parent state depicted in Fig. \ref{fig:cartoon}(b). We start again with a trilayer of free Dirac fermions near a $\nu=2\rightarrow1$ plateau transition,
\begin{align}
\mathcal{L}_{\mathrm{IQH}}&=\sum_{n=1}^3\left[\bar\Psi_n(i\slashed{D}_A-M)\Psi_n-\frac{3}{2}\frac{1}{4\pi}AdA\right]\,.
\end{align}
This theory is dual to a trilayer of Wilson-Fisher composite bosons, $\Phi_n$, coupled to fluctuating CS gauge fields, $\alpha_n$, \cite{Seiberg2016,Karch2016},
\begin{align}
\mathcal{L}_{\mathrm{IQH}}[A]\leftrightarrow\sum_n{\mathcal{L}}^\Phi_{n}[\Phi_n,\alpha_n,   A]&=\sum_n\left[|D_{\alpha_n}\Phi_n|^2-r|\Phi_n|^2-|\Phi_n|^4+\frac{1}{4\pi}\alpha_nd\alpha_n+\frac{1}{2\pi}Ad\alpha_n-\frac{1}{4\pi}AdA\right]\,,
\end{align}
where $-|\Phi|^4$ again denotes tuning such that the theory is at its Wilson-Fisher fixed point when $r=0$, and the phase diagrams of the two theories match if $\sgn(r)=-\sgn(M)$. 

We now attach a positive flux to the electric charges on layers $n=1$ and $3$ and a negative flux to those on layer $n=2$. This is implemented in a manifestly gauge invariant way by the following transformation on each layer's Lagrangian \cite{Kivelson1992,Witten2003},
\begin{align}
\label{eq: appendix flux attachment}
\mathcal{L}^\Phi_n[\Phi_n,\alpha_n,A]\rightarrow\mathcal{L}^\Phi_n[\Phi_n,\alpha_n,\gamma_n]+\frac{1}{2\pi}\gamma_nd\beta_n+\frac{(-1)^n}{4\pi}\beta_nd\beta_n+\frac{1}{2\pi}Ad\beta_n,
\end{align}
where $\beta_n,\gamma_n$ are new fluctuating $U(1)$ gauge fields. One can easily check that the electric charges in the gapped phases of this theory have had their statistics shifted by $\pm\pi$. Because the equation of motion for $\gamma_n$ is
\begin{align}
d(\alpha_n+\beta_n)=d\gamma_n\,,
\end{align}
$\gamma_n$ may be integrated out while preserving flux quantization. The resulting Lagrangian on each layer is
\begin{align}
\mathcal{L}^\Phi_n\equiv|D_{\alpha_n}\Phi_n|^2-r|\Phi_n|^2-|\Phi_n|^4+\frac{2}{4\pi}\alpha_nd\alpha_n+\frac{1}{4\pi}\left[1+(-1)^n\right]\frac{1}{4\pi}\beta_nd\beta_n+\frac{1}{2\pi}\alpha_nd\beta_n+\frac{1}{2\pi}Ad\beta_n\,,
\end{align}
where we have redefined $\mathcal{L}^\Phi_n$ to minimize the number of labels in use. On layers $n=1,3$, the CS term for $\beta_n$ vanishes. Integrating it out therefore Higgses $\alpha_n$ (in other words, sets $d\alpha_n=dA$), leaving a topologically trivial theory near a superconductor-insulator transition. On layer $n=2$, however, the CS term for $\beta_n$ has level 2, meaning that the gauge theory is topologically nontrivial and has the $K$-matrix of the Halperin (2,2,1) state. Explicitly, renaming $\alpha_2\equiv\alpha,\beta_2\equiv\beta$,
\begin{align}
\mathcal{L}^\Phi_{p}&=|D_{A}\Phi_{p}|^2-r|\Phi_{p}|^2-|\Phi_{p}|^4+\frac{2}{4\pi}AdA\,,\,p=1,3\,;\\
\mathcal{L}^\Phi_2&=|D_\alpha\Phi_2|^2-r|\Phi_2|^2-|\Phi_2|^4+\frac{2}{4\pi}\alpha d\alpha+\frac{2}{4\pi}\beta d\beta+\frac{1}{2\pi}\alpha d\beta+\frac{1}{2\pi}\beta dA\,.
\end{align}
The trilayer theory, $\sum_n\tilde{\mathcal{L}}^\Phi_n$, is the theory depicted in Fig. \ref{fig:cartoon}(b). 

We now check that these theories are dual to theories of composite vortices, which on clustering yield the Fibonacci state. Applying the duality used in Eq. \eqref{eq: bosonic parent} of the main text along with the transformation flux attachment transformation in Eq. \eqref{eq: appendix flux attachment}, the dual theories of composite vortices are
\begin{align}
\tilde{\mathcal{L}}^\phi_n\leftrightarrow\tilde{\mathcal{L}}_p^\phi=\,&|D_a\phi_p|^2-\tilde{r}|\phi_n|^2-|\phi_n|^4+\frac{1}{4\pi}\Tr\left[a_nda_n-\frac{2i}{3}a^3_n\right]\nonumber\\
&+\frac{1}{2\pi}\gamma_nd\Tr[a_n]+\frac{1}{2\pi}\gamma_nd\beta_n+\frac{(-1)^n}{4\pi}\beta_nd\beta_n+\frac{1}{2\pi}Ad\beta_n\,,
\end{align}
where again $a_n$ are $U(2)$ gauge fields. In this case, both $\gamma_n$ and $\beta_n$ can be safely integrated out without running afoul of flux quantization: integrating out $\gamma_n$ implements a constraint on (i.e. Higgses) $\beta_n$, $d\beta_n=-d\Tr[a_n]$. The resulting theories involve $U(2)_{1,-1}$ gauge theories on layers $n=1,3$, which is topologically trivial \cite{Hsin2016}, and a $U(2)_{1,3}$ theory on the $n=2$ layer,
\begin{align}
\label{eq: appendix composite vortices}
\tilde{\mathcal{L}}^\phi_n=|D_a\phi_p|^2-\tilde{r}|\phi_n|^2-|\phi_n|^4+\frac{1}{4\pi}\Tr\left[a_nda_n-\frac{2i}{3}a^3_n\right]+(-1)^n\left[\frac{1}{4\pi}\Tr[a_n]d\Tr[a_n]+\frac{1}{2\pi}Ad\Tr[a_n]\right]\,.
\end{align}
As in the discussion in the main text, we are free to invoke charge conjugation symmetry to flip the sign of the BF term on layer $n=2$ relative to those on layers $1,3$. From here, it is straightforward to see that a nonzero expectation value for the clustering order parameter, $\langle\Sigma_{mn}\rangle=\langle\phi^\dagger_m\phi_n\rangle\neq0$, sets $a_1=a_2=a_3\equiv a$ and produces the Fibonacci $U(2)_{3,1}$ TQFT,     
\begin{align}
\mathcal{L}_{\mathrm{Fib}}&=\frac{3}{4\pi}\Tr\left[a_nda_n-\frac{2i}{3}a^3_n\right]-\frac{1}{4\pi}\Tr[a]\,d\Tr[a]+\frac{1}{2\pi}Ad\Tr[a]\,.
\end{align}
Integrating out the $\phi_n$ fields but leaving the Fibonacci order parameter thus leads to the final Landau-Ginzburg theory obtained in the main text,
\begin{align}
\mathcal{L}=&\sum_{m,n}\Tr\left[|\partial\Sigma_{mn}-ia_m\Sigma_{mn}+i\Sigma_{mn} a_n|^2\right]-V_r[\Sigma]\nonumber\\&+\sum_n\left[\mathcal{L}_{\mathrm{CS}}[a_n]
%&+\sum_n\mathcal{L}_{\mathrm{CS}}[a_n]+b_\mu j^\mu_{\mathrm{top}}+\frac{1}{4\pi}(b+A)d(b+A)\,,
+(-1)^n\left(\frac{1}{4\pi}\Tr[a_n]d\Tr[a_n]+\frac{1}{2\pi}Ad\Tr[a_n]\right)\right]\,.
\end{align}
\subsection{Representation of the Fibonacci order in terms of $U(2)_{3,1}$}

%Equivalence of $(G_2)_1$ and $U(2)_{3,1}$ as TQFTs}

In this section, we demonstrate explicitly that $U(2)_{3,1} = [SU(2)_3 \times U(1)_2]/\mathbb{Z}_2$ possesses the same anyon content as that of $(G_2)_1$, namely, just the Fibonacci anyon. 
%To do so requires understanding how to implement the $\mathbb{Z}_2$ quotient on the $SU(2)_3 \times U(1)_2$ topological order. As described in Sec. \ref{sec: SM-CS-conventions}, at the level of the field theory, this corresponds to combining the $SU(2)$ and $U(1)$ gauge fields into a single $U(2)$ gauge field. 
There are multiple ways to describe the process of enforcing the $\mathbb{Z}_2$ quotient in the definition of $U(2)_{3,1}$. From the perspective of the anyon content of the theories, this quotient amounts to condensing \cite{Bais2009} a bosonic anyon in the $SU(2)_3 \times U(1)_2$ product theory with $\mathbb{Z}_2$ fusion rules and either $0$ or $\pi$ braiding statistics with all other anyons. The condensed anyon %-- the generator of a discrete $\mathbb{Z}_2$ symmetry -- 
is then identified as a local quasiparticle, and so all anyons with which it braids nontrivially are projected out.
In order to identify the anyon to be condensed, let us remind ourselves of the anyon content of the $SU(2)_3$ and $U(1)_2$ factors:
\begin{align}
		U(1)_2 &: \qquad 1, \, s \\
	SU(2)_3 &: \qquad [0], \,  [1/2], \,  [1], \, [3/2].
\end{align}
Here, $s$ is the semion, which has topological spin $h_s = 1/4$ and satisfies the fusion rule $s \times s = 1$. We have labelled the anyons of $SU(2)_3$ by the representation of $SU(2)$ under which they transform. They are all self-dual, satisfying the fusion rules
\begin{align}
	[0] \times [0] &= [0] \\
	[1/2] \times [1/2] &= [0] +[1] \\
	[1] \times [1] &= [0] +[1] \\
	[3/2] \times [3/2] &= [0].
\end{align}
From this, we see that $[1]$, which has spin $h_{[1]} = 2/5$, is the Fibonacci. The only Abelian anyon is $[3/2]$, which has spin $h_{[3/2]} = 3/4$, trivial braiding with $[1]$, and non-trivial braiding with $[1/2]$. We immediately see that, in the product theory, $[3/2] s$ is an Abelian anyon with spin unity. %It is the generator of the $\mathbb{Z}_2$ symmetry and on condensing it, 
On condensing this anyon, all anyons aside from the Fibonacci will become confined, yielding the desired $(G_2)_1$ Fibonacci topological order. 

\subsection{Constructing the Electron Operators}% Constructing the ${(G_2})_1$ current algebra}

As stated in the main text, the electron operators used in constructing the Fibonacci wave function must be selected from the generators of the $(G_2)_1$ current algebra. We present the technical details of this process here. 
The $(G_2)_1$ current algebra has fourteen generators, twelve of which are labeled by the roots of $G_2$. In order to obtain explicit expressions for these operators, we make use of the duality between $(G_2)_1$ and $U(2)_{3,1} = [SU(2)_3 \times U(1)_2]/\mathbb{Z}_2$, which will allow us to write the generators in terms of operators in the $SU(2)_3$ and $U(1)_2$ conformal field theories (CFTs).
%We make use of the identity
%\begin{align}
%	(G_2)_1 \cong U(2)_{3,1} = \frac{SU(2)_3 \times U(1)_2}{\mathbb{Z}_2}.
%\end{align}
%In order to determine the generators of the current algebra of this CFT, we first spell out the structures of the $U(1)_2$ and $SU(2)_3$ factors and then work out the $\mathbb{Z}_2$ quotient. % As we shall see below, arranging the generators on the basis of the $G_2$ root system will make more transparent our identification of the electron operators. 

The $U(1)_2$ factor is described by a chiral boson, $\varphi$, with compactification radius $R=1$. It supports a single anyon, the semion, represented by the vertex operator
\begin{align}
    s(z) \equiv e^{i\varphi(z) / \sqrt{2}},
\end{align}
which has scaling dimension $\Delta_s = 1/4$. The operators $s^2 = e^{i\sqrt{2}\varphi}$ and $\bar{s}^2 = e^{-i\sqrt{2}\varphi}$ generate the $U(1)_2$ chiral algebra, and so correspond to local excitations.

%Turning to $SU(2)_3$, we first note that the primary fields of $SU(2)_k$ are organized by the representation of $SU(2)_k$ under which they transform: $[j]$, $j=0,1/2,\dots,k/2$. These operators satisfy the fusion rules
%\begin{align}
%	[j_1] \times [j_2] = \sum_{j=|j_1-j_2|}^{\min(j_1+j_2,k-j_1-j_2)} [j].
%\end{align}
%Hence, $SU(2)_3$ has primary fields falling into 
As for $SU(2)_3$, its primary fields, like the anyons in the corresponding TQFT, fall into four topological sectors labelled by the $SU(2)$ representation under which they transform: $[j]$, $j=0,1/2,1,3/2$. In order to write down explicit forms of these fields and the current operators, we make use of the %equality
%\begin{align}
%	SU(2)_3 = \frac{\mathrm{Parafermion}_3 \times U(1)_6}{\mathbb{Z}_3}, \label{eq: SU(2)_3}
%\end{align}
fact that the operators of $SU(2)_3$ can be expressed in terms of products of operators in the $k=3$ parafermion and $U(1)_6$ CFTs, the former of which we will write as \para. %{\raman (cite di Francesceo et. al.? Also need better notation than \para for the parafermion CFT)}. 
The $U(1)_6$ CFT is described by a chiral boson $\phi$ at radius $R=1$, with primary fields
\begin{align}
    a^l(z) \equiv e^{il\phi / \sqrt{6}}, \quad l=0,\dots , 5
\end{align}
These fields have scaling dimensions $\Delta_l = l^2/12$, from which we see that the field $a^6$ represents a local excitation. 
The primary fields of the \para CFT and their scaling dimensions are given in Table \ref{tab:para-fields} while their fusion rules are given in Table \ref{tab:para-fusion}. 
The raising and lowering operators of the $SU(2)_3$ algebra are given by the operators, %$SU(2)_3$ algebra is then generated by the current operators
\begin{align}
	\psi_1 a^2 = \psi_1 e^{i\sqrt{2/3} \phi}, \quad \psi_1^\dagger \bar{a}^2 =  \psi_2 e^{-i\sqrt{2/3} \phi}. %,
\end{align}
%with the former representing an electron and the latter a hole.

\begin{table}[h!]
 \begin{tabular}{|| c | c | c | c | c | c | c ||} 
 \hline
  & $1$ & $\psi_1$ & $\psi_2$ & $\sigma_1$ & $\sigma_2$ & $\epsilon$ \\
 \hline\hline
 $\Delta$ & 0 & $2/3$ & $2/3$ & $1/15$ & $1/15$ & $2/5$ \\ 
 \hline
\end{tabular}
\caption{Scaling dimensions of the \para primary fields.} \label{tab:para-fields}
\end{table}

\begin{table}[h!]
 \begin{tabular}{|| c | c | c | c | c | c ||} 
 \hline
  $\times$ & $\psi_1$ & $\psi_2$ & $\sigma_1$ & $\sigma_2$ & $\epsilon$ \\
 \hline\hline
 $\psi_1$   & $\psi_2$   & & & & \\ 
 $\psi_2$   & $1$        & $\psi_1$   & & & \\ 
 $\sigma_1$ & $\epsilon$ & $\sigma_2$ & $\sigma_2+\psi_1$ &  & \\ 
 $\sigma_2$ & $\sigma_1$ & $\epsilon$ & $1+\epsilon$      & $\sigma_1+\psi_2$ & \\ 
 $\epsilon$ & $\sigma_2$ & $\sigma_1$ & $\sigma_1+\psi_2$ & $\sigma_2+\psi_1$ & $1+\epsilon$ \\ 
 \hline
\end{tabular}
\caption{Fusion rules of \para.} \label{tab:para-fusion}
\end{table}

Now, in order to obtain the $(G_2)_1$ algebra from $SU(2)_3 \times U(1)_2$, we must perform the $\mathbb{Z}_2$ quotient. As in the TQFT description, this corresponds to condensing operators in the %As noted in the main text, this corresponds to condensing the operators in the
\begin{align}
	\left[ \frac{3}{2} \right] s %\{ \bar{s}, s \}
\end{align}
topological sectors. In the language of CFT, this ``condensation" means that the operators in these topological sectors will be identified as generators of the $[SU(2)_3 \times U(1)_2]/\mathbb{Z}_2$ (equivalently, $(G_2)_1$) CFT. Explicitly, the operators
\begin{align}
	\bar{a}^3, \quad \psi_1 \bar{a}, \quad \psi_2 a, \quad a^3
\end{align}
are all in the $[3/2]$ sector, and so are topologically equivalent. Indeed, each is related to the other by fusion with the $SU(2)_3$ generators, forming an $SU(2)_3$ quartet. Hence, performing the $\mathbb{Z}_2$ quotient means condensing the operators
\begin{align}
	\begin{split}
	\bar{a}^3 s, \quad \psi_1 \bar{a} s, \quad \psi_2 a s, \quad a^3 s \\
	\bar{a}^3 \bar{s}, \quad \psi_1 \bar{a} \bar{s}, \quad \psi_2 a \bar{s}, \quad a^3 \bar{s}
	\end{split}
\end{align}
%and the corresponding operators with the exchange $s \leftrightarrow \bar{s}$. 
This set of operators, combined with the generators of $SU(2)_3$ and $U(1)_2$ constitute the twelve generators of $(G_2)_1$ labelled by its roots \cite{Schoutens2016}. %Explicitly, we have \cite{Schoutens2016}
%\begin{align}
%	\begin{split}
%		 \psi_1 a^2,& \qquad \psi_1 \bar{a}s, \qquad \psi_1 \bar{a} \bar{s}, \\
%		 \psi_2 \bar{a}^2 ,& \qquad \psi_2 a \bar{s} , \qquad \psi_2 as, \\
%		 a^3 \bar{s},& \qquad a^3 s , \qquad s^2, \\
%		 \bar{s}^2 ,& \qquad \bar{a}^3 \bar{s} , \qquad \bar{a}^3 s.
%	\end{split}
%\end{align}
%Note that all of these operators have scaling dimensions of unity. %The remaining two generators are the Cartan generators and are given by linear combinations of $\partial \phi$ and $\partial \varphi$. 

\begin{figure}[!htb]
    \includegraphics[width=0.35\textwidth]{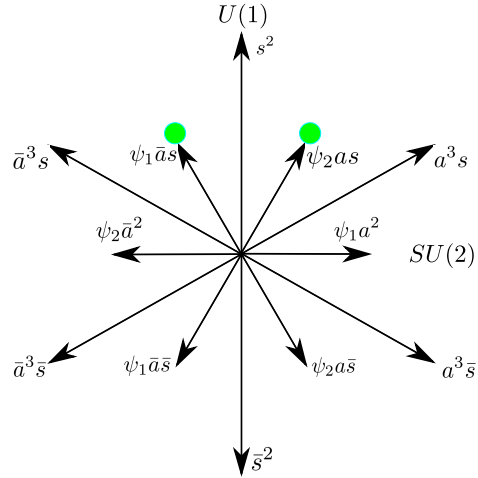}
  \caption{ Root system of $G_2$ labelled by the corresponding $(G_2)_1$ current generators. The green circles indicate the operators we identify as the electron operators.} \label{fig:g2-algebra}
\end{figure}

Fig. \ref{fig:g2-algebra} depicts the $G_2$ root system labeled by the corresponding current generators. One can check that vector addition of the roots matches up with fusion of the corresponding current operators. Note also that the generators naturally organize themselves in terms of their transformation properties under $SU(2)$ and $U(1)$. The vertical coordinate of the root corresponds to the $U(1)$ charge and the horizontal coordinate to the $SU(2)$ spin. %In particular, $\psi_1 a^2$ and $\psi_2 \bar{a}^2$ correspond to the raising and lowering operators for the $SU(2)$ spin (being the generators of $SU(2)_3$). 

It now remains to determine which generators we should identify as the physical electrons. In the spirit of Refs. \cite{Ardonne1999,Ardonne2002}, we expect that we must choose two electron operators, by virtue of the fact that the root system is two-dimensional. %that the electron operators should correspond to short roots (with the long roots being interpreted as condensate operators). 
The electrons should have the same positive charge, suggesting we should restrict ourselves to the upper half-plane of the root system. As described in the main text, we expect the Fibonacci wave function to describe a two-flavor system, and so the electron operators should have opposite $SU(2)$ spin. We thus claim that
\begin{align}
    \begin{split}
	\Psi_\uparrow &\equiv \psi_2 as =\psi_2 e^{i\phi / \sqrt{6} + i\varphi/\sqrt{2}} , \\	\Psi_\downarrow &\equiv \psi_1 \bar{a}s = \psi_2 e^{-i\phi / \sqrt{6} + i\varphi/\sqrt{2}}
	\end{split}\label{eq: SM-electron-operator}
\end{align}
are the appropriate electron operators. 

We note that operators $\bar{a}^3s$ and $a^3s$ also satisfy our two criteria for charge and spin. In fact, $\Psi_\uparrow$ and $\Psi_\downarrow$ form an $SU(2)$ quartet with $\bar{a}^3s$ and $a^3s$ (as can be seen from Fig. \ref{fig:g2-algebra}), and so one may reasonably ask whether the latter two operators constitute equally valid choices for the electron operator. As it turns out, the wave function obtained from $\bar{a}^3s$ and $a^3s$ describes an \emph{Abelian} state, as we demonstrate in the following section. This suggests, \textit{a posteriori}, that $\Psi_{\uparrow / \downarrow}$ are the correct electron operators needed to obtain a wave function describing the non-Abelian Fibonacci state.

%\begin{widetext}
\subsection{Derivation of the Fibonacci Wave Function}
In this section, we present a computation of the explicit form of the Fibonacci wave function provided in the main text. With the choice of electron operators given in Eq. \eqref{eq: SM-electron-operator}, we can express the wave function as
\begin{align}
	\Psi(\{ z_i, w_i \} ) = \langle \prod_{i=1}^N \Psi_\uparrow(z_i) \Psi_\downarrow (w_i) O_{bg} \rangle = \langle \prod_{i=1}^N \psi_2 a s (z_i) \psi_1 \bar{a} s (w_i) O_{bg} \rangle ,
\end{align}
where $z_i$ and $w_i$ label the positions of the up and down spins (spin is used as a stand-in for some flavor index). Here $O_{bg}$ is a background charge operator that ensures the correlator of the $s$ fields is non-vanishing and yields the usual Gaussian factor on the plane \cite{Difrancesco-1997}. Note that such an operator for the $a$ fields is \emph{not} necessary, since there are an equal number of $a$ and $\bar{a}$ fields, ensuring their charge neutrality condition is satisfied. Physically, this is a consequence of the fact that it is the $U(1)_2$ sector and hence the $s$ fields which are charged under the external electromagnetic field. We thus obtain (dropping the usual overall Gaussian factor),
\begin{align}
	\Psi(\{ z_i, w_i \} ) &= \langle \prod_{i=1}^N \psi_2(z_i) \psi_1(w_i) \rangle \langle \prod_{i=1}^N e^{i \frac{1}{\sqrt{6}}\phi (z_i)} e^{-i \frac{1}{\sqrt{6}}\phi (w_i)} \rangle \langle \prod_{i=1}^N e^{i \frac{1}{\sqrt{2}}\varphi (z_i)} e^{i \frac{1}{\sqrt{2}}\varphi (w_i)} O_{bg} \rangle \\
	&= \langle \prod_{i=1}^N \psi_2(z_i) \psi_1(w_i) \rangle \prod_{i,j} (z_i - w_i)^{1/3} \prod_{i<j} (z_i - z_j)^{2/3} \prod_{i<j} (w_i - w_j)^{2/3}.
\end{align}
In order to evaluate the remaining correlator, we can use the parafermion operator product expansions (OPEs),
\begin{align}
    %\begin{split}
	\notag\psi_1(z) \psi_1(z') &\sim (z-z')^{-2/3} \psi_2 (z') + \dots \, \text{(likewise for } 1\leftrightarrow 2) \\
	\psi_1(z) \psi_2(z') &\sim (z-z')^{-4/3} + \dots \, . \label{eq: SM-parafermion-ope}
	%\end{split} 
\end{align}
to effectively point-split the $\psi_2$ operators into products of $\psi_1$ operators:
\begin{align}
	\langle \prod_{i=1}^N \psi_1(z_i^1) \psi_1(z_i^2) \psi_1(w_i) \rangle = \langle \prod_{i=1}^N (z_i^1-z_i^2)^{-2/3} \psi_2(z_i^2) \psi_1(w_i) \rangle + \dots \, ,
\end{align}
where, here and in the following, the limit $z_i^1 \to z_i^2$ is taken implicitly. The ellipsis represent less singular terms in the $\psi_1\times\psi_1$ OPE which vanish in this limit, allowing us to isolate the desired parafermion correlator when we take $z_i^1 = z_i^2 \equiv z_i$ at the end of the computation.

Now, the correlator of $\psi_1$ fields is precisely given in terms of the Read-Rezayi (RR) wave functions:
\begin{align}
	\langle \prod_{i=1}^N \psi_1(z_i^1) \psi_1(z_i^2) \psi_1(w_i) \rangle = \Psi_{RR}^{k=3}(\{ z_i^1 , z_i^2 , w_i \}) \Psi_{LJ}^{-2/3}(\{ z_1^1 , z_i^2 , w_i \}).
\end{align}
Here, $\Psi_{RR}^{k=3}$ and $\Psi_{LJ}(\{z_i\}) = \prod_{i<j} (z_i-z_j)$ are the $\nu=3/2$ bosonic RR (taking $k=3$ and $M=0$ in the notation of Ref. \cite{Read1999}) and Landau-Jastrow wave functions, respectively. Hence,
\begin{align}
	\langle \prod_{i=1}^N \psi_2(z_i^2) \psi_1(w_i) \rangle + \ldots &=  \Psi_{RR}^{k=3}(\{ z_i^1 , z_i^2 , w_i \}) \Psi_{LJ}^{-2/3}(\{ z_1^1 , z_i^2 , w_i \}) \prod_{i=1}^N (z_i^1-z_i^2)^{2/3} \\
	\begin{split}
	&= \Psi_{RR}^{k=3}(\{ z_i^1 , z_i^2 , w_i \}) \prod_{i<j} (z_i^1 - z_j^1)^{-2/3} (z_i^2 - z_j^2)^{-2/3} (w_i - w_j)^{-2/3} \\
	& \qquad\qquad \times \prod_{i\neq j}(z_i^1 - z_j^2)^{-2/3} \prod_{i,j} (z_i^1 - w_j)^{-2/3} (z_i^2 - w_j)^{-2/3} .
	\end{split}
\end{align}
We can now safely set $z_i^1 = z_i^2 \equiv z_i$, in which case the terms contained in the ellipsis vanish identically. Combining terms and ignoring unimportant overall phase factors, we obtain
\begin{align}
    \Psi(\{z_i, w_i \}) &= \Psi_{RR}^{k=3}(\{ z_i , z_i , w_i \})\prod_{i<j} (z_i - z_j)^{-2}\prod_{i,j} (z_i - w_j)^{-1}
\end{align}
as our Fibonacci wave function. Here, $\Psi_{RR}^{k=3}(\{ z_i , z_i , w_i \})$ is the bosonic $\nu=3/2$ RR wave function for $3N$ particles, with the coordinates of $N$ pairs of these particles set equal to one another. As noted in the main text, the asymmetry in $z_i$ and $w_i$ is a consequence of having point-split the $\psi_2$ parafermions as opposed to the $\psi_1$ parafermions. Had we instead point-split the $\psi_1$ parafermions into products of $\psi_2$ parafermions, we would have obtained the above expression with $z_i$ and $w_i$ exchanged. Since the expressions obtained via these two different point-splitting procedures must necessarily be equal, we can write down the wave function in a manifestly symmetric way by taking their average:
\begin{align}
	\Psi(\{ z_i, w_i \} )&= \frac{1}{2} \left( \frac{\Psi_{RR}^{k=3}(\{ z_i , z_i , w_i\} )}{ \prod_{i<j} (z_i - z_j)^{2}} + \frac{\Psi_{RR}^{k=3}(\{ z_i , w_i , w_i\} )}{ \prod_{i<j} (w_i - w_j)^{2}} \right) \prod_{i,j} (z_i - w_j)^{-1}.
\end{align}

Finally, we return to the remark regarding the choice of electron operators made at the end of the preceding section. Had we instead attempted to construct a wave function using $\Psi_{\uparrow\uparrow} = a^3 s$ and $\Psi_{\downarrow\downarrow} = \bar{a}^3 s$ as the electron operators, we would have obtained
\begin{align}
    \tilde{\Psi}(\{ z_i, w_i \} ) = \langle \prod_{i=1}^N \Psi_{\uparrow\uparrow}(z_i) \Psi_{\downarrow\downarrow} (w_i) O_{bg} \rangle = \langle \prod_{i=1}^N e^{i \sqrt{\frac{3}{2}}\phi (z_i)} e^{-i \sqrt{\frac{3}{2}}\phi (w_i)} \rangle \langle \prod_{i=1}^N e^{i \frac{1}{\sqrt{2}}\varphi (z_i)} e^{i \frac{1}{\sqrt{2}}\varphi (w_i)} O_{bg} \rangle\, .
\end{align}
The correlators of vertex operators can be straightforwardly evaluated to obtain
\begin{align}
    \tilde{\Psi}(\{ z_i, w_i \} ) = \prod_{i<j} (z_i - z_j)^2 (w_i - w_j)^2 \prod_{i,j} (z_i - w_j)^{-1},
\end{align}
which describes the \emph{Abelian} Halperin $(2,2,-1)$ state, again at filling $\nu=2$. This gives us some confidence that $\Psi(\{z_i,w_i\})$ correctly describes the Fibonacci state. 

%\end{widetext}

\end{document}